# Review of Different Privacy Preserving Techniques in PPDP

Jalpesh Vasa[#], Panthini Modi[*]
[#]*Assistant Professor, Dept. of Information Technology, CSPIT, CHARUSAT, Changa, India*
[*]*Student, Dept. of Information Technology, CSPIT, CHARUSAT, Changa, India*

**Abstract -** *Big data is a term used for a very large data sets that have many difficulties in storing and processing the data. Analysis this much amount of data will lead to information loss. The main goal of this paper is to share data in a way that privacy is preserved while information loss is kept at least. Data that include Government agencies, University details and Medical history etc., are very necessary for an organization to do analysis and predict trends and patterns, but it may prevent the data owner from sharing the data because of privacy regulations [1]. By doing an analysis of several algorithms of Anonymization such as k-anonymity, l-diversity and t-closeness, one can achieve privacy at minimum loss. Admitting these techniques has some limitations. We need to maintain trade-off between privacy and information loss. We introduce a novel approach called Differential Privacy.*

**Keywords -** *Big data, Privacy, k-anonymity, l-diversity, t-closeness, Differential Privacy, Correlation, Privacy-Preserving Data Publishing (PPDP).*

## I. INTRODUCTION

Big data means a very large amount of data. Big data has shape hope to the world. As big data became a major problem in the modern world, it has created a huge hazard in preserving one's privacy. Privacy is a major issue in the current scenario in the world when one wants to make use of data that involves individual's information. This is the problem of an organization such as Shopping, bank or insurance company wants to release data to the public, requires privacy as data in original form, may contain sensitive information and publishing such data will violate individual's privacy. Privacy-Preserving Data Publishing (PPDP) provides some methods and tools for publishing useful information while big data privacy is preserved. Each record has a number ofattributes. In PPDP, attributes are classified asPersonal Information Identifiers (PII), Quasi Identifiers (QI), Sensitive Attributes (SA), and Non-sensitive Attributes [2].

### A. *Personal Information Identifiers (PII)*
Information which can directly identify individual's identity, such as their name, id or any other information which is linked to an individual.

### B. *Quasi Identifier (QI)*
QI can combine with other entity to re-identify and become personal information, such as age.

### C. *Sensitive Attributes (SA)*
It is personal information which is not supposed to disclose, such as salary, disease.

### D. *Non-sensitive Attributes*
Information which can be published or disclosed.

This paper illustrates k-anonymity, l-diversity and t-closeness with its pros and cons. To prevent disclosing one's identity, Samarati and Sweeny introduced k-anonymity as the property that each record is indistinguishable with at least k-1 other records with respect to the quasi-identifier [4]. It has 2 methods: Generalization and suppression which is done by using correlation coefficient. K-anonymity solves the solution of identity disclosure but does not prevent attribute disclosure problem as well as Homogeneity attack. In order to avoid the limitation of k-anonymity, the notion of l-diversity has been proposed but it is also insufficient to prevent attribute disclosure problem. Moreover, it has the limitation of similarity attack. So particular paper introduces a novel approach that is t-closeness. It is calculated by the Earth Mover Distance (EMD). But t-closeness got some drawback. So, Cynthia Dwork in 2006 [3] has popularized the term **Differential Privacy**. Which wipe out almost all drawback of Anonymization techniques. Dwork has shown that it is impossible to publish information from a private statistical database without revealing some amount of private informationand that the entire database can be revealed by publishing the results of a surprisingly small number of queries. In this section, we are comparing Anonymization techniques with Differential Privacy through some measurements to evaluate the techniques. Evaluation criteria such as Performance, Utility, Information loss etc. will be measured. We can provide privacy in several phases.

**Big data privacy process has 3 phases:**
- Data Generation
- Data Storage
- Data Processing

There are some techniques of Anonymization to provide privacy in these phases.

## II. K-ANONYMITY

To prevent record linkage attack, Samarati andSweeney popularized the concept of k-anonymity. The basic idea is to anonymize the data such that each individual cannot be distinguished from a group of other individuals in the data [5]. In other words, A release of data is said to have the **k-anonymity** property if the information for each person contained in the release cannot be distinguished from at least **k**-1 individuals whose information also appear in the release.

**Table I. Original Table**

| Age | Gender | Country | Religion | Diseases |
|---|---|---|---|---|
| 55 | Male | China | Hindu | Viral_infection |
| 25 | Female | Germany | Christian | TB |
| 42 | Female | China | Muslim | Heart_Attack |
| 49 | Male | Dubai | Budhh | Heart_Attack |

There are two techniques by which k-anonymity can be achieved.

### A. Generalization
In which each record of attributesisgeneralized. For example, generalizing Age attribute in which one person's age is 25, then it may be generalized by '20<age<30' or '<=30' or '>20' etc.

### B. Suppression
In this technique, certain values of n attributes are replaced by an asterisk '*' sign. This is called 'Blocking' the value. Blocking all the character or a number of each value may lead to data loss. So, some or half of the values are blocked using '*'.

The result of above methods shows that these operation causes a considerable amount of information loss because higher the generalization hierarchy more information loss will be there [1]. Also, suppression causes the elimination of values. This paper eliminates the drawbacks by using **Correlation Coefficient**, which reduces the loss of important data. Moreover, we do not have to generalize and suppress the value manually. It is done by the correlation coefficient.

## III. CORRELATION COEFFICIENT

To know which Quasi Identifier will do generalization and which QID will do suppression, there is a process called 'Attribute Selection'. This process returns the correlation coefficient of two QID; for instance, Age and Religion. If the answer of the correlation coefficient is:

Greater than zero, then both the attributes are positively correlated. That means if the value of one attribute increase, correspondingly the value of another attribute will also increase.
- Less than zero, then both the attributes are negatively correlated. Which means increases one value of the attribute will decrease the other value of the attribute.
- Zero, then no correlation between both of the QID.

Which attribute has higher correlation will be undergoingGeneralization and which are having lower correlation will be subjected to Suppression.To calculate the correlation coefficient between Age and Religion.

Consider, Age X
Religion Y

$$Correl(X,Y) = \frac{\sum(x-x')(y-y')}{\sqrt{\sum(x-x')^2(y-y')^2}}$$

**Table II. Anonymized Table**

| Age | Gender | Country | Religion | Diseases |
|---|---|---|---|---|
| >=50 | Male | China | * | Viral infection |
| <=25 | Female | Germany | * | TB |
| 25<Age<50 | Female | China | * | Heart Attack |
| 25<Age<50 | Male | Dubai | * | Heart Attack |

Although, k-anonymity has some limitations. Such as Homogeneity attack, Background knowledge attack, attribute disclosure problem etc. To overcome the limitations of k-anonymity, a further technique that has been introduced over here is **l-diversity**. In the next section, the paperillustrates l-diversity with proper example.

## IV. L-DIVERSITY

An equivalence class is said to have an l-diversity ifthere are at least l "well-presented" values for the sensitive attribute [4]. It is an extension of k-anonymity which diminishes the granularity of data representation utilizing methods including Generalization and Suppression in a way that any given record map onto at least k different records in the data.

**Table III Diverse Table**

| Age | Gender | Country | Religion | Diseases |
|---|---|---|---|---|
| >=50 | Male | China | * | Viral infection |
| <=25 | Female | Germany | * | TB |
| 25<Age<50 | Female | China | * | Heart stroke |
| 25<Age<50 | Male | Dubai | * | Heart Attack |

In short, PPDP technique; l-diversity protects againsthomogeneity attack but it still does not come up with the solution of attribute disclosure problem. If the values for one (or several) confidential attributes (s) in all records are the same, then the intruder learns the values of that (those) attribute(s) for the target individual T [6].A further limitation of l-diversity is; it is difficult to achieve and it does not cover up the similarity attack. After l-diversity's failure, the next technique has been introducing is; t-closeness. We will discuss this following.

## V. T-CLOSENESS

T-closeness is another extension of k-anonymitywhich try to solve the attribute disclosure problem.An equivalence class is said to have t-closeness if the distance between the distribution of a sensitive attribute in this class and the distribution of the attribute in the whole table is no more than a threshold t. A table is said to have t-closeness if all equivalence classes have t-closeness [4].

Distribution of two sensitive attributes is supposed to keep close to each other. But the problem is to measure the distance between two probabilistic distributions. However, there are so many alternatives to calculate distance. Such asVariational distance and Kullback-Leibler (KL) distance etc. but they do not reflect the semantic distance among values.So, we have a ground distance which is defined by any pair of values. We want distance between the twoprobabilistic distribution to be dependent upon ground distance. Earth Mover's Distance (EMD), which satisfies this need. More generally, EMD is the total work divided by total flow. EMD is computed as the minimum transportation cost by moving distribution mass between each other.So, it depends on how much mass is moved and how far mass is moved [7].

### A. EMD for t-closeness

By calculating EMD between two distributions, one can solve the transportation problem such as min-cost flow. The EMD is based on the minimal amount of work needed to transform one distribution to another by moving distribution mass between each other [8]. It is widely used in content-based image retrieval to compute distancesbetween the colour histograms of two digital images.

### B. EMD for Numerical Attributes

The natural order can be used to measure the distance. There is a clear ground distance. The ground distance is computed as ordered distance. The EMD between P and Q can be computed as [9]:

$$EMD(P, Q) = \frac{1}{r-1} \sum_{i=1}^{r} \left| \sum_{j=1}^{i} (pi - qi) \right|$$

Let P and Q are probability distributions.

### C. EMD for Categorical Attributes

Order does not exist. In categorical attributes, there is no relation between attribute values. So, it is better to set ground distance as 1 between any two different attribute values. The EMD of categorical attributes[9]:

$$EMD(P = Q) = \frac{1}{2} \sum_{i=1}^{r} |pi - qi|$$

But, only hiding some information does not assure the protection of individuals privacy.So,the latest notion that has been introducedis Differential Privacy, which is announced in next section.

## VI. DIFFERENTIAL PRIVACY

This is the most recent technique used to provideprivacy. Differential privacy is a process to add randomness into the data, which provides the solution of the problem. The term was introducing by Cynthia Dwork in 2006. Privacy emerges from the falsifiability of individual responses.All the above techniques of privacy preservation have a big concern to maintain a tradeoff between privacy and utility. Differential Privacy investigates stability between these two parameters; privacy and utility, in which data will remain useful, as well as disclosure risk, must be limited.

For example, assume that we have database $D_1$ ofHIV test, which contains the name of the patient. The record is a pair of (Name, X) where X is a Boolean value indicating person has HIV or not.

**Table IV. Differential Privacy**

| Name | Has HIV (X) |
|---|---|
| Eli | 1 |
| Justin | 1 |
| Bob | 0 |
| Lisa | 1 |
| Alice | 0 |

Consider adversary wants to find whetherAlice has HIV or not. He knows Alice resides on which row of the database $D_1$.Suppose adversary is only allowed to use a query that returns the partial sum of the X column. In order to check Alice's HIV status, adversary calculates$Q_5(D_1)$ and $Q_4(D_1)$. That is, $Q_5(D_1)$ = 2 and $Q_4(D_1)$ = 3. Now, if adversary wants to find the result of Alice, partial sum of X i.e.; $Q_1$ to $Q_4$ becomes 3. So, adversary comes to know that Alice value is 0. This is how individual's information can be revealed without knowing information about a particular person. Differential Privacy provides a solution for this problem by adding some noise to the dataset. If we change the value of Alice by 1 in dataset $D_2$ and make it available publicly then adversary would not be able to know personal information.

Since differential privacy is the probabilistic concept,

let P be the probability,
A → randomized algorithm,
$D_1$ and $D_2$ → datasets,
Ɛ → positive real number.

$$P[A(D_1) \in S] \leq e^{\varepsilon} * P[A(D_2) \in S]$$

According to the formula, the difference between both the datasets must be less or equal to $e^{\varepsilon}$.

It does not hold the solution. This can be achieved by adding random Laplace noise to each value of the dataset $D_1$ to achieve differential privacy. This approach is called Laplace Mechanism.

It does not hold the solution. This can be achieved by adding random Laplace noise to each value of the dataset $D_1$ to achieve differential privacy. This approach is called Laplace Mechanism.

Anothermechanism, the Laplace mechanism is used in differential privacy to add Laplace noise with low sensitivity. Suppose we have a dataset $D_1$ of the healthrecord. We want to release a number of Heart diseases patient in $D_1$ which is 1000. By applying above formula:

$$P[A(D_1=1000) \in S] \leq e^{\varepsilon} * P[A(D_2=1000) \in S]$$

**Table V. Pros and Cons of Algorithms**

| Privacy Model | Pros | Cons |
|---|---|---|
| k-anonymity | Evaluating systems that release information such that released information limits what can be revealed about the properties of entities that are to be protected. Protect against identity disclosure problem. | Homogeneity-attack background knowledge attack Does not provide a solutionto attribute disclosure problem. |
| l-diversity | The adversary has background knowledge of the information. | Difficult to achieve. Insufficient to prevent attribute disclosure problem. Similarity attack |
| t-closeness | An equivalence class is said to have t-closeness if the distance between the distribution of a sensitive attribute in this class and the distribution of the attribute in the whole table is no more than a threshold t. | T-closeness requires that the distribution of a sensitive attribute in any equivalence class is close to the distribution of a sensitive attribute in the overall table. Does not deal with identity disclosure. |
| Ɛ-Differential Privacy | Provides solution of all these algorithms by adding Laplace noise. | The balance between utility and privacy still challenges. |

## VII. CONCLUSION AND FUTURE WORK

In this paper, firstly we have started with theintroduction of big data and general information about the algorithms. Then,we haveexamined the different privacy-preserving techniques one by one, discussing whether existing techniques are enough to process the big data. All the techniques were briefly argued with an example.K-anonymity protects against identity disclosure, it does not provide sufficient protection against attribute disclosure. The notion of l-diversity attempts to solve homogeneity attack and background knowledge attack but it cannot resolve the problem of attribute disclosure So we have proposed a novel privacy notion called t-closeness. We use Earth Mover's Distance for t-closeness but it is certainly not perfect. Recent technology which is used to protect privacy is Differential privacy. It provides strong privacy; even adversary has arbitrary external knowledge. Moreover, we discussed pros and cons of each technique. Traditional as well as recent both techniques were reviewed in this paper. In future, we can compare these anonymization techniques withdifferential privacy using various evaluation criteria. Also, we can use multiple sensitive attributes.